\documentclass[sigconf, screen]{acmart}
\usepackage[]{hyperref}
\usepackage{color}
\usepackage{tabularray}
\usepackage{tabularx}
\usepackage{tikz}
\usepackage{multirow}

\usepackage{geometry}
\usepackage{boxedminipage}
\usepackage{setspace}

\usepackage{xurl}

\newcommand{\phold}[1]{{\color{black}#1}}

\tikzstyle{every number}=[draw=black] 

\newcommand{\ignore}[1]{}
\newcommand*\whitecircled[1]{\tikz[baseline=(char.base)]{
            \node[shape=circle,inner sep=1pt,draw] (char) {\bfseries\footnotesize #1};}}

\AtBeginDocument{%
  }



\copyrightyear{2026}
\acmYear{2026}
\setcopyright{cc}
\setcctype{by}
\acmConference[ICS '26]{2026 International Conference on Supercomputing}{July 06--09, 2026}{Belfast, United Kingdom}
\acmBooktitle{2026 International Conference on Supercomputing (ICS '26), July 06--09, 2026, Belfast, United Kingdom}
\acmDOI{10.1145/3797905.3800521}
\acmISBN{979-8-4007-2522-7/2026/07}

\begin{document}

\title{IBEX: Internal Bandwidth‑Efficient Compression Architecture for Scalable CXL Memory Expansion}

\author{Younghoon Ko}
\orcid{0009-0007-0685-6407}
\affiliation{%
  \department{Electrical and Computer Engineering \& ISRC}
  \institution{Seoul National University}
  \city{Seoul}
  \country{South Korea}
}
\email{ykhoon3810@snu.ac.kr}

\author{Hyemin Park}
\orcid{0009-0008-5347-7685}
\affiliation{%
  \department{Electrical and Computer Engineering \& ISRC}
  \institution{Seoul National University}
  \city{Seoul}
  \country{South Korea}
}
\email{hmpark98200@capp.snu.ac.kr}

\author{Hyuk-Jae Lee}
\orcid{0000-0001-8895-9117}
\affiliation{%
  \department{Electrical and Computer Engineering \& ISRC}
  \institution{Seoul National University}
  \city{Seoul}
  \country{South Korea}
}
\email{hjlee@capp.snu.ac.kr}

\author{Hyokeun Lee}
\orcid{0000-0002-0824-6238}
\affiliation{%
  \department{Electrical Engineering and Computer Science}
  \institution{DGIST}
  \city{Daegu}
  \country{South Korea}
}
\email{hyokeun.lee.phd@gmail.com}

\begin{abstract}
As the memory channel count is confined by physical dimensions, memory expanders appear to be a promising approach to extending memory capacity and channels by augmenting the existing I/O interface (e.g., PCIe) with memory-semantic protocols like CXL. 
Unfortunately, the physical constraints of a computing system restrict scalable capacity expansion with memory expanders. 

In this work, we propose a block-level compression scheme for modern memory expanders, \textit{IBEX}, to achieve larger effective memory capacity. Given the performance overhead associated with block-level compression algorithms (e.g., LZ77), IBEX employs a promotion-based approach: only cold data is compressed, whereas hot data remains uncompressed. Our key innovation is internal bandwidth-efficient block management that precisely identifies cold pages with minimal metadata access overhead.
Still, the promotion-based approach poses several performance-related challenges at the design level. Therefore, we also propose a shadowed promotion scheme that temporarily postpones the deallocation of promoted data, thereby mitigating the performance penalty incurred by demotion (i.e., recompression). Furthermore, we optimize our compression scheme by compacting metadata and co-locating multiple target blocks for efficient bandwidth utilization.
Consequently, IBEX achieves an average of 1.28$\times$--1.40$\times$ speedups compared to the state-of-the-art promotion-based block-level approaches.
We open-source IBEX at \url{https://github.com/relacslab/ibex-ics26}. 
\end{abstract}

\maketitle

\section{Introduction}
\label{sec:intro}

\textit{Memory capacity wall} has become one of the severe concerns due to the ever-increasing gap between the working set sizes of emerging workloads and the growth in memory density. 
For example, service applications like social networks or navigation are based on graph analytics that comprise trillions of nodes (e.g., users or locations) and edges (e.g., connections or paths) \cite{graph1, graph2, graph3, graph4}. AI assistants based on large language models or recommendation models generate hundreds of gigabytes of data during runtime \cite{batch1,batch2,batch3,batch4,oasis}. 
On the platform side, the pin counts on a processor are on the verge of saturation \cite{kevin-fam-isca09,clio-asplos22,sdm-pact23}, deterring further increases in the number of memory modules. Thus, high-speed, high-capacity memory becomes pressing to accommodate such scenarios.

To overcome the capacity wall, companies in the industry have developed various technologies. For example, high-density memory cells like non-volatile memory (NVM) have been announced \cite{mram,pcm,sttram}; however, NVM has been struggling with power, thermal, and reliability issues for decades \cite{pcm-energy,pcm-survey,stt-thermal,pcm-wde}, resulting in limited density scaling and availability. 

Recently, Compute Express Link (CXL)---an off-chip interconnection technology---has emerged as a key technology for scalable memory expansion. As stated in several previous works \cite{dtl-isca23,directCXL-atc22,tpp-asplos23,pond-asplos23,sdm-pact23,salus-hpca24}, \textit{CXL memory expander (or simply memory expander)} facilitates the adoption of various memory technologies (e.g., DRAM or NAND), as CXL preserves abstract load/store semantics with minimal modifications and performance overheads at the software level \cite{hellobytes}. Furthermore, the CXL protocol is defined over the high-speed physical interface for I/O devices (e.g.,  64GB/s for PCIe Gen 5.0 $\times$16). Therefore, memory expansion that supports memory semantics via non-DIMM interfaces can not only augment capacity and bandwidth but also provide a broader product portfolio.

However, capacity extension with memory expanders still faces various physical constraints in a practical computing system. 
Conventionally, a computer system is assembled on a standardized motherboard (e.g., ATX, micro-ATX), on which components like CPU, memory, and I/O devices reside together. Accordingly, standardizing the physical dimensions of these components is highly recommended by defining \textit{form factors} to incorporate more components in a space-efficient manner. 
Recently, \textit{Enterprise and Datacenter Storage Form Factor (EDSFF)}, which was originally devised as SSD's "frame box" \cite{snia-ssdformfactors}, has also been adapted for CXL memory \cite{micron-cxl,hynix-cxl} to achieve scalable and robust accommodation. Since the form factor constrains the volume of such devices, the number of memory devices inside a memory expander would be limited.
Additionally, a small number of PCIe slots on a standardized motherboard also limits the number of memory expanders installed in a system. For instance, high-end servers typically offers around eight (e.g., Dell PowerEdge XE7745 \cite{dell-poweredge}) or ten (e.g., Dell PowerEdge XE9680 \cite{dell-israel-1}) PCIe slots; however, many of these slots are typically occupied by GPUs and accelerators for AI workloads that necessitate high computing power \cite{dell-israel-1}, leaving only a handful available for memory expanders. 
Consequently, the limitations in slot count and per-device capacity strongly motivate the incorporation of \textit{memory compression} in memory expanders as a more scalable solution. Recently, the Open Compute Project (OCP), a consortium comprising more than 50 semiconductor companies, announced the specification of hyperscale CXL memory, which includes the \textit{block-level compression} feature to maximize effective capacity \cite{ocpspec}.

Nevertheless, introducing block-level compression into a memory expander entails new challenges. 
First, a scalable data caching scheme must be devised to reduce latency while achieving the high compression ratio of block-level compression. Block-level compression, which symbolizes patterns within a larger block unit (typically configured as a few kilobytes), yields a much larger effective capacity compared to line-level compression that compresses data at the cacheline granularity \cite{fpc,bdi,cpack,bpc,lcp,buri,cmh,rmc,compresso,buddy}. However, block-level compression incurs significant latency overhead upon decompression (e.g., 64 cycles for LZ77 \cite{mxt}) due to its high algorithmic complexity and the potential need for fetching larger data blocks. 
As an alternative, the overhead might be mitigated by using large SRAM-based caches \ignore{(i.e., for either touched data or compression metadata)} within memory expanders; however, the form factor disallows vendors from freely spending their budgets on hardware. 
Therefore, in general, block-level compression adopts \textit{data promotion (or migration)} to store touched data by provisioning a dedicated uncompressed memory region \cite{mxt,dmc,tmcc}, yielding lower access latency. 
Second, compression-related traffic would be bottlenecked by the \textit{limited internal bandwidth} of memory expanders, as their form factor restricts the number of available channels. For instance, promotion-based compression could incur frequent migrations for memory-intensive workloads. Moreover, promotion-based compression typically introduces compression metadata to check data status and locate the compressed data stored in unaligned formats, further burdening internal bandwidth.

Previous promotion-based block-level compression works targeting traditional main memory have optimized the performance from various aspects \cite{mxt,dmc,tmcc, dylect, dmu}; however, these works still have limitations in hyperscale systems that incorporate CXL memory. 
For example, additional address translation is required in compression to translate physical addresses recognized by the OS into device addresses of compressed or uncompressed regions. Panwar et al. \cite{tmcc} embeds the translation information in page table entries to hide its latency. However, this approach would require non-trivial modifications to the CPU microarchitecture, potentially compromising availability in hyperscale data centers, where the system uptime is critical.
Kim et al. \cite{dmc} proposed a hardware-oriented approach that decouples translation information from page table entries as \textit{metadata} to significantly reduce such design complexity. This work also separately manages frequently accessed data (including its consecutive neighbors) by unifying the compressed format with a simpler compression scheme (e.g., line-level), whereas infrequently accessed ones are managed with block-level schemes to ensure high compression ratios. Hence, the access overhead of decoupled metadata is alleviated as well. Still, this heterogeneous approach requires high-bandwidth memory (e.g., HMC) to migrate several pages at a time, whereas the internal bandwidth of CXL is very limited. 
Panwar et al. \cite{dylect} seeks huge-page-like translation performance by introducing a short version of metadata for frequently accessed pages through set associative page placement. To maintain full addressability, cold pages continue to use conventional full-length metadata, and the system dynamically switches between short and normal entries based on page access frequency. However, this design requires maintaining two separate metadata tables, necessitating the controller to consult both during translation. The dual table lookup renders a design that is inefficient for internal bandwidth-constrained CXL memory.

\noindent\textbf{Our solution:} 
To address these limitations, we propose IBEX, an internal bandwidth-efficient compression architecture tailored for practical memory expanders. IBEX introduces a bandwidth-efficient compressed block management mechanism that precisely and efficiently identifies cold blocks for demotion, thereby minimizing memory management overhead. The key idea is to employ a hardware-friendly variant of the second-chance algorithm to select cold blocks, along with a lazy update mechanism for efficient tracking of page references; hence, only minimal traffic overhead is induced. Moreover, the identification of inactive blocks is achieved through the page activity region managed internally within the device, transparent to the OS. Thus, our proposed approach considers both the limited internal bandwidth and the system availability of data centers. 
Additionally, given that dynamic data access patterns in hyperscale systems may trigger frequent migrations and metadata accesses, we propose several optimizations to reduce internal bandwidth consumption associated with these operations.
First, we propose a novel \textit{shadowed promotion} scheme that puts off the deallocation of to-be-promoted data to minimize the penalty of future demotion.
Additionally, we provide a design option to allow the \textit{co-location} of multiple blocks to be managed using single metadata instead of provisioning one metadata for one compressed block. 
Finally, we tailor and compact the compression metadata to achieve only a single metadata access, even when shadowed promotion and co-location schemes are applied by flexibly \textit{repurposing} the metadata based on compression status.

To evaluate our IBEX, we extend the SST simulator \cite{sst} to develop a CXL-based computing system. According to our evaluation, IBEX achieves the speedups of 1.28$\times$--1.40$\times$ compared to promotion-based block-level compression schemes \cite{tmcc,dylect}.

\section{Background}
\label{sec:background}
\subsection{CXL and Memory Expander}
CXL is an open-standard interconnect protocol defined over a high-speed physical interface to enable high-bandwidth communication between different components. Since the CXL version 3.x leverages PCIe Gen 6.0 as its physical interface \cite{cxl-spec}, CXL can reach up to 128GB/s of bandwidth if 16 lanes are available on the target device, comparable to that of dual-channel DDR5 modules (2$\times$8B$\times$6.4Gbps=102.4GB/s). 
Furthermore, the CXL allows the processors to access devices using load/store semantics (i.e., load/store instructions) with a finer-grained transfer unit, ranging from 64B to 256B, by augmenting the root complex in processors; hence, CXL devices exhibit much lower latency than conventional block devices that need to transfer a block of data and go through the block I/O layer in operating systems (OS) \cite{tpp-asplos23}. 
Owing to CXL's high bandwidth, low latency, and transparency, a CXL type-3 device (or simply \textit{CXL memory} or \textit{memory expander}) facilitates memory expansion connected via available I/O interfaces rather than solely adhering to traditional DIMM interfaces; this allows systems to process larger datasets and memory-intensive applications simultaneously, showing solid benefits in scalability.

\subsection{Data Compression Algorithms} \label{sec:compression}
Data compression transforms the data into smaller-sized chunks by symbolizing repeated patterns (or \textit{symbols}) along with their occurrence frequency. Thus, compression can enhance either the effective bandwidth utilization or capacity or both. Depending on the operating data size, there are primarily two types of compression: line-level compression and block-level compression. 
Line-level compression compresses cacheline-sized data (typically 64B) by identifying repeated patterns as reference symbols and representing the data with reference symbols and their occurrence frequencies. The reference symbol can be established based on various policies \cite{cpack,bdi,bpc,fpc,sc2}, such as repeated patterns, the number of zeros, etc. Due to its algorithmic simplicity, line-level compression is widely applied to main memory compression, yielding a moderate compression ratio and low decompression latency. 

Unlike line-level compression, block-level compression compresses larger data units---\textit{block} ($\geq$1KB). To achieve a higher compression ratio than the line level, block-level compression typically establishes a \textit{dictionary} that moves along the block being compressed, like a sliding window \cite{lz4, lz7, zstd, beezip}. Such a sliding dictionary can dynamically identify new or existing symbols within the block; hence, it is more likely to find repeated patterns in larger data, yielding a higher compression ratio than line-level compression. 
However, the benefit of the compression ratio comes at the expense of the latency overheads of matching patterns in a large block. Therefore, block compression has been widely used in latency-tolerant cases, such as storage devices.

\subsection{Design Considerations of Data Compression in a Memory Subsystem} \label{sec:background-main_memory}
In addition to compression and decompression, applying compression to practical memory systems raises design concerns. First, \textit{compression metadata} is necessary to check the stored position and compression status (i.e., compressed, uncompressed, or all zeros). In a linear address space, multiple compressed data are stored in compact and unaligned forms (e.g., three compressed data are stored across two cachelines). Therefore, the compression engine must be able to locate the position of each compressed data; hence, the metadata contains the addresses of multiple compressed data chunks to locate them once the data is accessed. 

Second, the implementation of compression metadata is another design concern in practical memory systems. A system can be aware of the status of all data in possibly two ways: OS-managed and OS-transparent approaches. In the OS-managed approach, the OS not only stores additional information (e.g., compression status and compressed sizes) in each page table entry but also manages variably sized pages using complex data structures, like traditional reservation-based paging \cite{reserv-pg-osdi02}. Moreover, the coherence management of such compressed but shared pages necessitates OS modifications.
To avoid such complex modifications, the OS-transparent approach has become ubiquitous in modern memory compression \cite{compresso,buri}. In this approach, metadata is \textit{separately} defined beyond the normal data region and is invisible to software (i.e., physically indexed). Hence, this approach allows the OS to manage the page-granularity address as usual. Specifically, after a virtual address is translated to the OS physical address (OSPA at page granularity), the dedicated hardware logic \textit{translates the OSPA to the memory physical address (MPA)} after referencing the metadata.

\section{Motivation}
\label{sec:motivation}

\subsection{Physical Constraints of CXL Memory} \label{sec:motiv-formfactor}
Besides the limited number of PCIe slots on systems, the physical volume constraints of CXL memory (i.e., \textit{form factor}) hinder the free expansion of memory capacity. According to the estimation from market products, we found that 32.8\% of the memory expander's area is currently occupied by DRAM devices\footnote{The ratio number is obtained from a 96GB E3.S 2T CXL memory that includes 32 DDR5 $\times$4 24Gb devices, announced in 2022 \cite{hynix-cxl} (i.e., eight devices per channel on each rank), where the areas of each DDR5 device and E3.S form factor are \phold{11mm$\times$8mm} \cite{micron-d5-24Gb} and \phold{76mm$\times$112.75mm} \cite{snia-ssdformfactors,snia-e3formfactors}, respectively.}. Since the rest of the area is potentially occupied by other modules (e.g., CXL controller, ECC parity devices \cite{micron-cxl,hynix-cxl}), the space for accommodating DRAM devices is scarce on the CXL memory as shown in \cite{micron-cxl-inside}. 
Expanding the form factor could be a viable solution; however, most enterprise-grade servers still use SSI-EEB boards, which have remained unchanged since the late 1990s’ E-ATX standard for cost-efficiency. 
Thus, block-level compression presents an opportunity to go beyond the form factor limit. 

In large-scale computing systems (e.g., hyperscale data centers), providing a scalable and ready-for-use solution without compromising system availability is critical. Adding a memory compression feature to either an OS or CPU-side memory controller requires the complete redevelopment of computing systems (e.g., kernel updates, CPU redesigns), resulting in unexpected costs associated with system management from the service provider's perspective. Hence, a vendor-side solution (or device-level compression) would be more preferable for scalably expanding memory size. As concrete evidence, OCP's hyperscale CXL memory specification \cite{ocpspec} includes block-level compression as a basic feature in CXL memory.

\begin{figure}[h]
\centering
\includegraphics[scale=0.75]{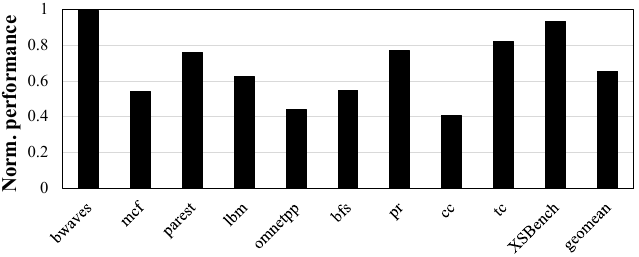}
\caption{Normalized performance of compressed CXL memory using dual-channel relative to ideal bandwidth case.} \label{fig:bandwidth_degradation}
\end{figure}

\subsection{Challenges of Block-Level Compression} \label{sec:motiv-challenge} 
As we understand why block-level compression is needed inside CXL memory, we now discuss the practical challenges of applying block-level compression in the context of CXL memory. 
Specifically, the \textit{limited internal bandwidth} poses a challenge in CXL memory; this is because, their channel count is bounded by their physical constraints, particularly the form factor. Block-level compression entails additional metadata accesses and amplified data transfers due to large block sizes. For instance, fetching a single cache line from a 2KB compressed block requires 32 memory accesses under a 64B granularity (common in most computing systems); modifying a single cache line similarly forces an update to the entire compressed block. While keeping frequently accessed blocks in uncompressed form can reduce redundant fetches and updates, this approach still incurs traffic overhead from managing and migrating the blocks within CXL memory.
Figure~\ref{fig:bandwidth_degradation} compares two compressed CXL memory configurations to quantify the impact of limited internal bandwidth: one constrained to dual DRAM channels (reflecting limited bandwidth) and another with effectively unlimited bandwidth but the same latency. As shown in this figure, the limited-bandwidth setup experienced an average performance degradation of 35\%, reaching 60\% in the worst case (\textit{cc}), highlighting the severity of bandwidth bottlenecks in compressed scenarios.

\begin{figure}[h]
\centering
\includegraphics[scale=0.75]{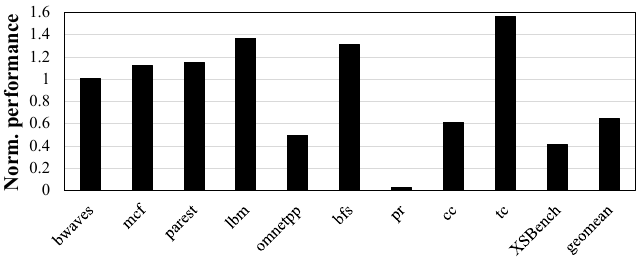}
\caption{Normalized performance of compressed CXL memory with naive SRAM relative to uncompressed CXL memory.}\label{fig:naive_sram}
\end{figure}

To address the internal bandwidth problem, a naive solution is to introduce an SRAM buffer that reduces redundant block fetches by reusing previously decompressed blocks. Figure~\ref{fig:naive_sram} depicts the normalized performance when a 16-way 8MB SRAM cache is configured to reuse recently decompressed blocks, where the performance is normalized to the baseline CXL configuration without memory compression. Most workloads show performance improvements, as cache hits are served without any memory accesses. However, memory-intensive workloads such as \textit{omnetpp}, \textit{pr}, \textit{cc}, and \textit{XSBench} show 76\% performance degradation. As a result, incorporating an SRAM cache for reusing recently decompressed blocks cannot avoid performance loss. Moreover, the form-factor problem of CXL memory limits the use of large SRAM, thereby constraining scalability. 

\section{Architecture}
\label{sec:design}
The primary goal of our paper is to significantly reduce the memory traffic overhead induced by memory compression, assuming a widely used and state-of-the-art \textit{promotion-based block-level compression technique} that stores hot or first-touched data in uncompressed form and cold data in compressed form \cite{tmcc}. To this end, we propose a bandwidth-efficient compressed block management that notably reduces internal bandwidth consumption related to block management. Before explaining the main idea of our proposed work, we first explain the baseline promotion-based block-level compression we assume for clarity. Subsequently, we would discuss the limitations of previous works that apply promotion-based compression in the context of practical CXL memory products. 

\subsection{Adapting Promotion-based Block-Level Compression to CXL Memory} \label{sec:baseline_promotion}
Similar to prior approaches \cite{mxt,dmc,tmcc}, CXL memory-based systems can mitigate the performance penalty incurred by block-level compression by adopting \textit{promotion-based data management}. Upon the first access of a compressed chunk, the compression engine decompresses it and stores it in the uncompressed region (i.e., \textit{promotion}) so that subsequent hits avoid costly decompression. Conversely, \textit{demotion} relies on recency hints to identify cold blocks, which are recompressed and written back in compressed form, freeing space for future promotions. Such a data management has been adopted by several block-level compression schemes and has proven effective in preserving capacity benefits while reducing performance penalties \cite{mxt,dmc,tmcc}.
%

\begin{figure}[h]
\centering
\includegraphics[scale=0.7]{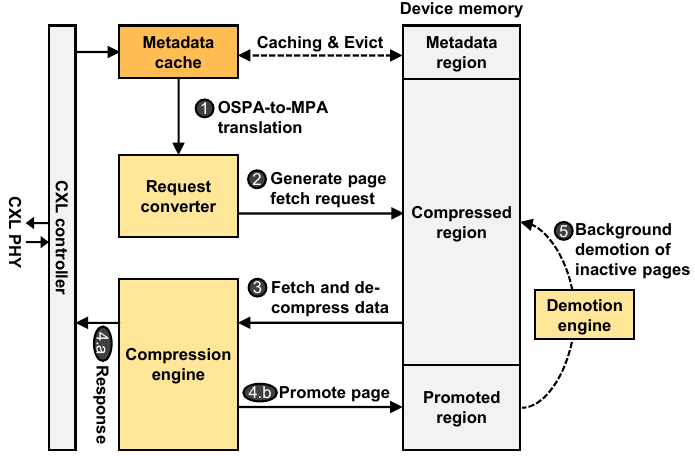}
\caption{The high-level procedures of promotion-based compression when accessing a compressed page.} \label{fig:key_process}
\end{figure}

Figure~\ref{fig:key_process} presents the promotion‑based compression that handles a memory request to compressed data in a CXL memory expander. Firstly, as a memory request arrives via the CXL interface (i.e., \textit{external request}), the OSPA-to-MPA translation is carried out using metadata in the \textit{metadata region} (Step \whitecircled{1}). 
The compression metadata holds the indirection information essential to translation and can be cached in a small-sized metadata cache (not the data cache) for faster translation. After translation, the \textit{request converter} generates one or multiple requests according to the data type recorded in the compression metadata. If the accessed data is compressed, the converter generates multiple read requests to fetch the entire compressed page stored in the compressed region (Step \whitecircled{2}). 
Subsequently, the compression engine decompresses the fetched data (Step \whitecircled{3}) to retrieve the target cacheline. The retrieved cacheline is then responded to the host-side (Step \whitecircled{4.a}). Meanwhile, the decompressed page is stored in the promoted region for future lower-latency access (Step \whitecircled{4.b}). 
When the promoted region runs out of space, the engine selects an inactive page according to the predefined policy and demotes it in the background (Step \whitecircled{5}).

\subsubsection{Memory Management} 
\label{sec:manage}
To implement promotion-based data management in CXL, three memory regions are required, namely the metadata region, the compressed region, and the promoted region. The metadata region can be provisioned at boot time by the driver or firmware and is visible to the compression engine; the other two regions store data according to their status. 
In the compressed region, the size of a compressed page varies with the bit patterns of data, where two possible allocation strategies are possible: (1) variable-size chunks and (2) fixed-size chunks.
Variable-size chunks allow each page to be allocated to a single, appropriately sized chunk. This approach typically leverages the features of \texttt{zsmalloc} \cite{zsmalloc}. Specifically, multiple 4KB pages are aggregated into a larger page (\textit{zspage}), which is then divided into smaller, demand-sized chunks \cite{tmcc,dylect,dmu}. Since page migrations leave holes within a zspage, it must track fine-grained zspage occupancy and periodically reclaim these fragments. Such additional traffic contradicts the limited internal bandwidth of CXL devices. Therefore, we adopt fixed-size chunks to simplify memory management.

To this end, we define chunks comprising compressed data as C-chunks, whereas the promoted data is composed of a larger unit, a P-chunk.
For instance, four 512B C-chunks (2KB of physical space) are allocated if a page is compressed to 2000B, whereas a promoted page consists of eight contiguous 512B chunks that form a single P-chunk.
Defining two chunk types requires using two linked lists to track free C-chunks and P-chunks belonging to the compressed region and the promoted region, respectively, as in a previous work \cite{buri}. In the compressed region, data is allocated with C-chunks to store compressed or incompressible data. 
On the other hand, the promoted region allocates data with P-chunks to store promoted data. As the number of free P-chunks falls below 256, demotion occurs in the background.
For both lists, a free chunk is selected by popping out the pointer stored in the head node. Also, the hardware necessitates two registers, each storing the pointer to the head of its corresponding free list.

\subsubsection{Translation Metadata} \label{sec:metadata}
\begin{figure}[h]
\centering
\includegraphics[scale=0.8]{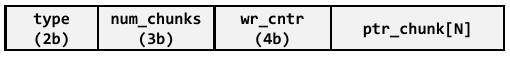}
\caption{Naive metadata entry format.} \label{fig:metadata}
\end{figure}

Since OSPA-to-MPA translation (Step \whitecircled{1} in Figure~\ref{fig:key_process}) is performed at page granularity, each metadata entry is responsible for a 4KB data block. Figure~\ref{fig:metadata} shows the naive metadata format based on such philosophy. The \texttt{type} field specifies the current data status that can fall into one of four categories: \textit{compressed} or \textit{promoted} for typical cases and \textit{zero} or \textit{incompressible} for special cases. Per-block metadata also holds the number of allocated chunks (\texttt{num\_chunks}) and pointers to the allocated chunks (\texttt{ptr\_chunk[i]} for i-th chunk). 

According to \texttt{ptr\_chunk[N]} field, selecting an appropriate chunk size involves a trade-off between compression ratio and metadata access overhead. Smaller chunks improve the compression ratio but increase the amount of pointer information stored within the metadata. To achieve the highest compression ratio while minimizing memory accesses for a metadata entry, we choose a chunk size of 512B---the smallest chunk size that allows metadata to fit within 64B (i.e., the access granularity of modern systems). We found that increasing the chunk size beyond 512B offers no additional memory access advantages and reduces the compression ratio. Conversely, choosing a smaller chunk size necessitates multiple memory accesses to retrieve the metadata while offering only marginal improvements in compression ratio (less than 7\% improvement for 256B chunks, according to our evaluation). Therefore, a chunk size of 512B represents a balanced design point of compression ratio and metadata access overheads.
The basic metadata needs 32-bit for each pointer of a 512B-chunk (i.e., \texttt{ptr\_chunk[i]}) to support a 41-bit physical address space (i.e., 2TB physical memory). Nevertheless, the assumption of this naive metadata entry format is reasonable and feasible to support the state-of-the-art server-grade CPU that can manage up to 4TB of memory per socket \cite{intel-cpu}, considering the total bit width of each metadata entry (i.e., 265b out of 512b for the 2TB case). We also assume that a metadata cache is employed to minimize latency, as in several works \cite{dmc, lcp, buri, cmh, rmc, compresso, tmcc}.

Fields in a metadata entry are used differently according to the data region where a data block is stored. For instance, the compressed pages are stored across \textit{one} to \textit{seven} 512B C-chunks, depending on the final compressed size. Thus, the size of a compressed block can be inferred using \texttt{num\_chunks}. 
Regarding the promoted pages, a single 4096B P-chunk is allocated to store the uncompressed data and occupy one \texttt{ptr\_chunk[i]} field in the metadata entry; hence, accessing a promoted page requires only a single read of an uncompressed cacheline. 
On the other hand, 8 C-chunks (not a P-chunk) and all eight \texttt{ptr\_chunk} fields must be occupied when block compression provides no benefit (i.e., incompressible page), as P-chunks are only reserved for highly compressible and frequently accessed pages. Although an incompressible page can become compressible after a few updates, trying compression for each update incurs significant traffic burdens on the compression engine. Therefore, \texttt{wr\_cntr} is defined in metadata to record the number of updates: when the number of updates exceeds the predefined threshold, the compression of the incompressible page is triggered as well as the \texttt{wr\_cntr} reset. According to our profiling, the threshold of 16 can potentially modify a quarter of a 4KB page. 
Lastly, a special case that must be handled is zero pages (i.e., pages consisting entirely of zeros). Similar to previous works \cite{rmc,fpc,zero-content}, zero page does not require any chunk.

\subsection{Limitations of Prior Promotion-based Block-Level Compression} \label{sec:limitation_promotion}
Several works have been proposed for promotion-based block-level compression to improve performance; however, they face limitations within the context of CXL memory.
\noindent\textbf{Reduction of translation overheads}: 
TMCC \cite{tmcc} reduces address translation overheads when accessing compressed and uncompressed data by \textit{embedding} translation metadata within a page table block (PTB). It first compresses every 64B PTB to spare free space. The freed space in compressed PTB is then reused to temporarily embed compression-translation entries (CTEs), each of which locates a compressed or uncompressed page, thereby eliminating translation overheads for usual accesses. However, this approach requires extensive cross-layer hardware modifications beyond the CXL device. For instance, the CPU-side L2 cache and memory controller must include PTB compression and decompression units. Furthermore, a speculative verification path is required to maintain correctness in the absence of OS-assisted synchronization of TLB-like CTE buffers across private L2 caches. Therefore, the design complexity issue would spread across CPU-side caches, memory controller, and MMU, resulting in a system-wide redesign and hence compromising the availability in hyperscale systems.
\noindent\textbf{Increasing metadata coverage}:
Decoupling metadata from the kernel data structure can significantly reduce design complexity at the expense of additional metadata access. Therefore, increasing the coverage of metadata entry is a key to reducing metadata accesses and improving the cacheability of metadata. For instance, DMC \cite{dmc} identifies frequently accessed data (i.e., hot data) and unifies the compression format of consecutive pages around that hot data. Specifically, the data, originally 32KB and compressed with block-level compression, is first migrated to the promoted region when accessed. Thereafter, the 32KB data is recompressed using the line-level compression scheme with a unified chunk size; hence, accessing any of the compressed data lines within the migrated block requires only one metadata entry access. However, this hardware-friendly, heterogeneous approach assumes HMC as the baseline system to migrate 32KB pages with high throughput, whereas our CXL memory has limited internal bandwidth compared to HMC.
Another work, DyLeCT \cite{dylect}, introduces a new memory region level that stores the most frequently accessed pages in addition to the promoted region. Unlike the fully associative promoted region, pages in this new region can only be placed within a designated set. Consequently, only a few bits of metadata are sufficient to locate a page at this level. Thereafter, DyLeCT aggregates multiple short metadata entries into a single cacheable metadata block, namely \textit{pre-gathered block}. However, this pre-gathered block does not significantly improve the metadata cache hit rate without OS support for huge pages. This is because, metadata is indexed using OSPAs, where spatial locality is not guaranteed across 4KB boundaries. Moreover, DyLeCT must probe both metadata tables (i.e., short and normal versions) on cache misses, resulting in additional traffic. This extra traffic may reduce efficiency in CXL memory, in which the internal bandwidth is limited.

\subsection{Our Goals} \label{sec:goals}
To address these limitations, we propose a comprehensive, bandwidth-efficient compressed block management mechanism that minimizes management overhead by profiling allocated pages within an OS-transparent \textit{page activity region}, namely \textit{IBEX} (Section~\ref{sec:page_management}). This management is performed internally within the CXL memory, eliminating the need for modifications to external hardware components (e.g., CPU) or the OS. 

\noindent\textbf{Design challenges:} 
Nevertheless, in hyperscale systems with CXL memory, workloads with large memory footprints exhibit dynamic data access patterns, posing some design challenges to IBEX.
First, demoting an incorrect block can lead to redundant traffic overhead due to repeated promotions of the same block. Therefore, our internal bandwidth-efficient block management scheme, IBEX, reduces traffic while ensuring accurate identification by smartly leveraging a page activity region that is transparent to the OS (Section~\ref{sec:page_management}). To avoid bandwidth utilization due to proactive updates of the activity region, we further employ a \textit{lazy update} scheme based on the status of page activity entries and leverage the eviction moment from the metadata cache.
Second, demotion may incur recompression as the demoted data must be compressed back to be stored in the compressed region, inevitably increasing traffic pressure on the internal memory channel. However, we observed that applications could have avoided such overheads with a novel \textit{shadowed promotion} (Section~\ref{sec:shadow}). 
Third, block-level compression leaves designers to select block sizes to balance the performance and compression ratio. Although a smaller block size provides an opportunity for higher performance, the number of metadata entries can grow. This is because, the metadata is allocated on a block basis, leading to more metadata accesses. Thus, we propose a methodology to smartly \textit{co-locate} multiple block information within a single metadata (Section~\ref{sec:colocate}). 
Lastly, block-level compression has various compression information (e.g., locating data, compression status). Therefore, carefully compacting the metadata is crucial to minimizing performance overheads due to metadata access (Section~\ref{sec:compact}). 

\subsection{Design of IBEX} \label{sec:page_management}

\begin{figure}[h]
\centering
\includegraphics[scale=0.75]{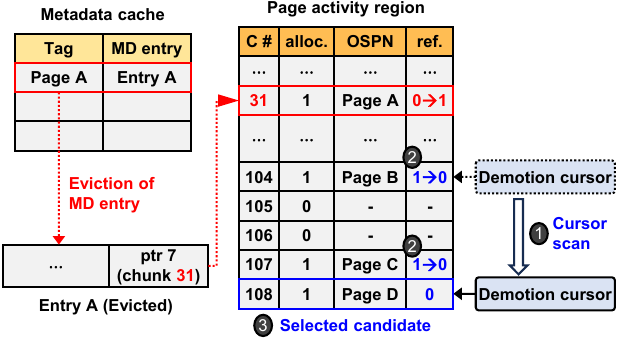}
\caption{IBEX handling reference bit update (red) and demotion candidate selection (blue).} \label{fig:block_management}
\end{figure}

\textit{Demotion policy} is a primary element of promotion-based block compression, as it identifies cold blocks to free space for newly promoted pages. IBEX's demotion policy is guided by two requirements imposed by CXL memory in hyperscale systems. First, it must be bandwidth-efficient to operate within the limited internal bandwidth of CXL memory. Although one possible approach is adopting the least-recently used (LRU) policy to select the demoting candidate, its doubly linked list implementation incurs several memory accesses per recency update, making it less feasible for bandwidth-constrained systems. Second, block management must remain OS-transparent to ensure deployability in hyperscale environments. To satisfy both requirements, we tailor a hardware-friendly variant of the second-chance algorithm and implement it using a dedicated region maintained within the device memory.

IBEX's demotion policy adds a dedicated page activity region alongside the three mandatory regions shown in Figure~\ref{fig:key_process}. Leveraging this additional region, our policy operates entirely in hardware, remaining transparent to the OS. To realize our policy, each 4B page activity entry should include \texttt{allocated} (1-bit), \texttt{OSPN} (30-bit), and \texttt{referenced} (1-bit). Such a format allows a single memory access to fetch \textit{16} activity entries (=64B/4B). \texttt{allocated} field indicates whether the P-chunk is assigned; \texttt{OSPN} stores the page number of OSPA; \texttt{referenced} represents whether the page has been \textit{recently} accessed.
Figure~\ref{fig:block_management} shows how the demotion engine scans the page activity region using the \textit{demotion cursor} (a hardware register) when the promoted region begins to run out of space (Step \whitecircled{1}). During the scanning, the \texttt{referenced} field is reset to 0 if the scanned activity entry has an allocated chunk (i.e., \texttt{allocated}=1) (Step \whitecircled{2}). Suppose the cursor touches a chunk with \texttt{referenced}=0 (i.e., not used for a while); in that case, the page assigned to that chunk is selected as a demotion candidate, triggering the demotion (Step \whitecircled{3}).
Since updating \texttt{referenced} on every page access would be too costly, we employ a lazy update scheme. Specifically, the \texttt{referenced} bit is updated only when the page's metadata is evicted from the metadata cache, as illustrated by the red text and arrows in Figure~\ref{fig:block_management}. This lazy update consolidates multiple reference bit updates, reducing bandwidth consumption.
Under the lazy update scheme, a hot page may be misidentified as a cold page if it remains in the metadata cache without eviction for an extended period. To prevent such pages from being demoted, the demotion engine probes the metadata cache to verify whether the page is indeed cold. If the page is found in the metadata cache, the cursor advances to the next entry without demotion.
Conversely, a cold page may be misidentified as a hot page if it persists in the metadata cache without eviction. However, we found that the probability of such cases is negligible. Considering a 16-way, 96KB LRU cache as in our evaluation (see Section~\ref{sec:eval-method}), the probability that an inactive entry remains in the cache falls below 0.01\% after 6,000 insertions. This implies that pages residing in the metadata cache are most likely within the recently accessed 24MB footprint. Consequently, it is safe to assume that these pages are effectively hot, given the hundreds of megabytes to gigabytes of the promoted region.

In rare cases when most pages are active (i.e., \texttt{referenced} is 1), the demotion engine may struggle to find a proper candidate, inducing undesired memory traffic. To limit worst-case bandwidth consumption, we combine the \textit{random} policy, which is known to offer reasonable performance in memory hierarchy \cite{random_replacement}; that is, one of the 16 fetched activity entries is randomly selected as the demotion candidate if no demotion candidate is found within a single 64B fetch. According to our evaluation, random selection rarely occurs (in just \textbf{0.6\%} of total selections). Therefore, our approach precisely identifies cold pages, minimizing unnecessary migrations.
Consequently, our policy achieves a 61\% reduction in memory traffic compared to a doubly linked list-based LRU implementation.

Please note that IBEX enhances the hardware architecture to achieve efficient compression control flow, rather than proposing a new compression algorithm. Hence, it provides flexible design choices among various block-level compression schemes, such as LZ4 \cite{lz4}, LZ77 \cite{lz7}, and Zstd \cite{zstd}. Since different schemes exhibit varying latency, we will show sensitivity to different compression latency values in Section~\ref{sec:eval-results}.

\subsection{Shadowed Promotion} \label{sec:shadow}
As mentioned in Section~\ref{sec:limitation_promotion}, demoting the page back to the compressed region requires \textit{recompression} of that page, potentially degrading the overall performance. For this reason, we propose novel \textit{shadowed promotion} by skipping the recompression process. 

\begin{figure}[h]
\centering
\includegraphics[scale=0.75]{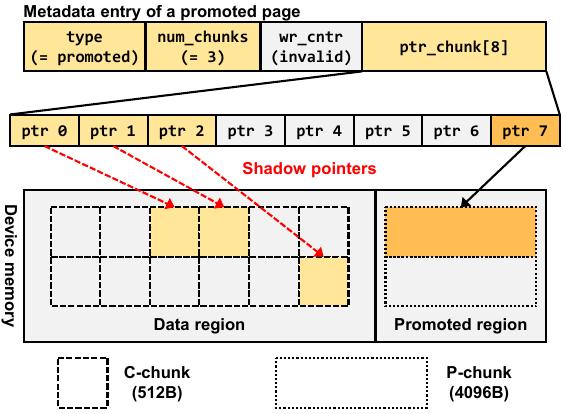}
\vspace{-0.5em}
\caption{Promoted page with shadowed promotion.} 
\label{fig:shadowed_promotion}
\end{figure}

As shown in Figure~\ref{fig:shadowed_promotion}, the philosophy behind the shadowed promotion is postponing the deallocation of C-chunks until the promoted page is updated. Preserving its original compressed copy instead of reclaiming it immediately, IBEX can later restore the page to its compressed form without recompression. Since practical workloads are mostly read-intensive, only a small fraction of promoted blocks require recompression. Notably, our evaluation shows around 62\% of total demotions in the baseline IBEX occur on unmodified data, implying that most promoted pages have been demoted back with unnecessary compression. 

Shadowed promotion does not require any additional metadata for tracking shadowed C-chunks. Since the maximum number of chunks mapped to a compressed page does not exceed 7, we can freely \textit{repurpose} the remaining one apart from seven allocated pointer fields (i.e., \texttt{ptr\_chunk[i]}) in metadata.
As a result, the preserved C-chunk pointers of the promoted page are only used as a backup to restore the page to its original state upon clean demotion (i.e., promoted data has never been updated). Specifically, the demotion process is simplified as deallocating the P-chunk and re-validating shadow pointers---it is simply achieved by updating the \texttt{type} field in metadata. 
Furthermore, the indication of a dirty page does not require an additional field, as the existence of a valid C-chunk pointer means that the page is still clean. 

Shadowed promotion duplicates data and hence would reduce actual memory savings; \textit{however}, this duplication is limited to the size of the promoted region, which is relatively small compared to the total device memory. For instance, assigning a 1GB promoted region within a 128GB device memory results in a maximum data duplication of 1GB, impacting the compression ratio by only 1\%. This cost is minimal compared to the potential capacity gain.

\subsection{Co-locating Information within Metadata} \label{sec:colocate}
Determining the block size in block-level compression involves a trade-off between performance and compression ratio. A larger block size can increase the compression ratio but also the latency of data decompression when accessing the compressed page. Therefore, we provide a possible configuration option facilitating the small block size of 1KB, as multiple previous works \cite{mxt,dmc} have shown that the 1KB block is the design point where the performance degradation is minimized with a reasonable compression ratio. 
However, a primary challenge behind this option is the increase of metadata resource overheads by 4$\times$, because block-level compression usually allocates a metadata entry for each block. 

\begin{figure}[h]
\centering
\includegraphics[scale=0.75]{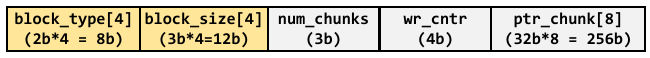}
\caption{Co-location-aware metadata format.} \label{fig:metadata-coloc}
\end{figure}

To address this challenge, IBEX decouples the compression granularity (block) from the page. By viewing a 4KB page as a set of four consecutive blocks, our proposed approach \textit{co-locates} all blocks' compression information in a single metadata entry while allowing individual compression/decompression for each 1KB block with low latency. 
Figure~\ref{fig:metadata-coloc} illustrates the enhanced metadata to support this co-location option. Our enhanced metadata additionally includes four pairs of [\texttt{block\_type}, \texttt{block\_sz}] (each pair for 1KB block), in place of \texttt{type} that has been defined in Figure~\ref{fig:metadata}. Specifically, the \texttt{block\_type[i]} indicates the compression status of an $i^{th}$ 1KB block, while the \texttt{block\_sz[i]} field indicates the size of the $i^{th}$ compressed block. 
Unlike prior block-level compression that assigns a metadata entry per 1KB block \cite{mxt, dmc}, our co-location facilitates each metadata to cover a 4KB range with a single metadata entry access, resulting in improved coverage and cacheability. Furthermore, it allows fine-grained promotion status using \texttt{block\_type} for each block, lowering the burden on internal bandwidth.

Having defined the enhanced metadata format, a naive approach to managing multiple compressed blocks is to store the compressed blocks while aligning each compressed block to multiples of 512B (i.e., the size of multiple C-chunks). For example, if two 1KB blocks are both compressed as 256B, these two compressed blocks are spread into two C-chunks. However, such a naive approach results in significant internal fragmentation within C-chunks, thereby resulting in low compression efficiency---In this example, 50\% of the space of a C-chunk is wasted.
As an alternative, we observe that introducing multiple pairs of [\texttt{block\_type}, \texttt{block\_sz}] can potentially allow a single C-chunk to accommodate several smaller compressed blocks. However, tracking the size of a compressed block at very fine granularity (say 1 byte) requires 10-bit for each \texttt{block\_sz[i]} and hence 40-bit in total for four 1KB blocks. 
To address this technical challenge, this paper aligns the smaller compressed block size to multiples of 128B. As a result, the 3-bit \texttt{block\_sz[i]} for each block (see Figure~\ref{fig:metadata-coloc}) yields values ranging from 0 to 7, each of which would become a multiplier (say $s$); hence, ($s$+1)$\times$128B indicates the actual compressed block size. Consequently, a C-chunk can maximally accommodate four compressed blocks that are compressed as 128B.

\subsection{Metadata Compaction} \label{sec:compact}
\begin{figure}[h]
\centering
\includegraphics[scale=0.74]{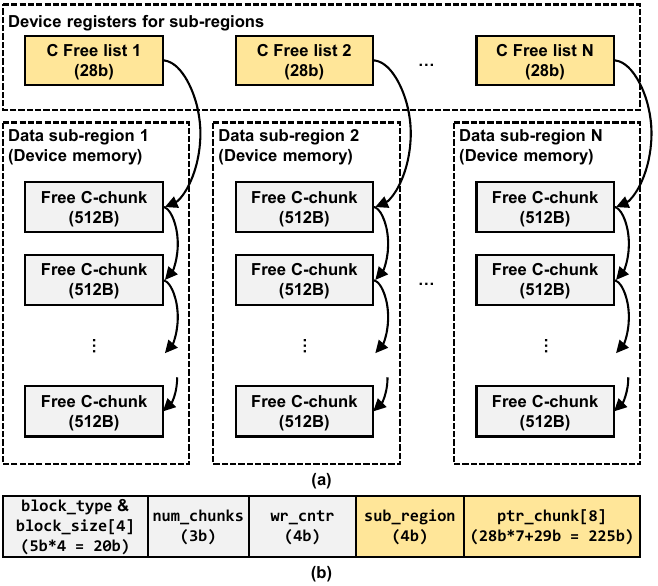}
\caption{Compacted (a) Free C-chunk list and (b) metadata.} \label{fig:metadata_compact}
\end{figure}

Although our enhanced compression metadata (283b in Figure~\ref{fig:metadata-coloc}) improves compression efficiency and cacheability, it may incur suboptimal performance and storage efficiency due to its unaligned data format. Specifically, nearly half of these entries cross the 64B boundary (i.e., access granularity in traditional computing systems) if these metadata entries are stored compactly; hence, two memory accesses are required to fetch entries that cross the 64B boundary. Based on this observation, aligning metadata entries to the 64B boundary while minimizing storage overhead is essential for minimizing additional memory access. 

Instead of simply aligning the existing metadata to a 64B boundary, we propose compacting the metadata format to minimize the associated memory access overheads without compromising expressiveness. 
To this end, we first divide the compression region into multiple sub-regions, where all C-chunks allocated to a single OSPA page must belong to the same sub-region. Figure~\ref{fig:metadata_compact} (a) shows modified linked lists of C-chunks; each linked list exists for each sub-region in addition to a header register. 
Based on our sub-region-based management, the C-chunk pointers of the same sub-region share some MSBs of a 41-bit physical address. Let's assume a sub-region of 128GB (i.e., 37-bit) as an example. C-chunks of a sub-region now share 4 MSBs (=41-37) in the metadata. Therefore, the bit width for each C-chunk pointer naturally becomes 28-bit (=37-9). 
Figure~\ref{fig:metadata_compact} (b) shows the final metadata format after compaction, which forms 32B per entry. Please note that the last pointer is compacted to 29 bits, as it needs enough bits to hold a pointer to a P-chunk. Consequently, a single 32B metadata entry of IBEX covers the translation of a 4KB OSPA range, necessitating only one memory access to fetch one or two metadata entries for 32B access granularity or 64B access granularity, respectively. 

\section{Evaluation Methodology} \label{sec:eval-method}

\begin{table} [h]
\caption{System configuration} \label{tab:sys-config}
\centering
\begin{tabular}{| p{0.2\linewidth} | p{0.7\linewidth} |}

\hline
\multicolumn{2}{|c|}{\textbf{Processor (4-core, \texttt{ariel})}} \\
\hline
\textbf{Core}               & 3.4Ghz, Out-of-Order, 4-issue/cycle \\ 
\hline
\textbf{L1 cache}           & Private, 8-way 64KB, LRU, 4-cycle        \\ 
\hline
\textbf{L2 cache}           & Private, 8-way 512KB, LRU, 10-cycle       \\ 
\hline
\textbf{L3 cache (LLC)}     & Shared,16-way 8MB, LRU, 20-cycle   \\ 
\hline
\hline
\multicolumn{2}{|c|}{\textbf{CXL memory expander}} \\
\hline
\multirow{2}{*}{\textbf{Interface}}     & Bandwidth: PCIe 5.0 $\times$8-lane \\ 
                                        & Latency: 70ns round-trip latency \\ 
\hline
\multirow{2}{*}{\textbf{Memory}}        & Dual channel, DDR5-5600, 128GB \\
                                        & tCL=40, tRCD=40, tRP=40 \\
\hline
\multirow{3}{*}{\textbf{Compression}}   & Metadata cache: 16-way 96KB, LRU, 4-cycle \\
                                        & Latency (compress/decompress) = 256-/64-cycle \\
                                        & Promoted region size = 512MB\\ 
\hline
\end{tabular}
\end{table}

We enhanced the Structural Simulation Toolkit (SST) \cite{sst} to simulate a CXL-based system. 
Table~\ref{tab:sys-config} shows the configuration of the processor and CXL memory. On the memory side, CXL's flit-based communication is modeled using \texttt{hr\_router}. The communication latency is set to 70ns, which complies with the target latency value outlined in \cite{cxl-spec}. 
The compression engine is implemented in the memory controller module using \texttt{timingDRAM}. According to previous work \cite{mxt}, the throughputs of compression and decompression engines are 4B/clock and 16B/clock, respectively, assuming 1KB block size; hence, we configure both latency values accordingly as shown in the table. 
On our platform, we also modify \texttt{ariel} (i.e., the processor model) to hook file-related function calls that read and write actual data, as the compressed data sizes determine the number of memory requests required to fetch the compressed data. Considering practicality in real scenarios, a random page allocation policy at the OS level (i.e., allocation of OSPAs) is adopted.

Since compression in CXL memory targets \textit{larger capacity} \cite{ocpspec, streamliningCXL}, we compare IBEX primarily against promotion-based block-level compression schemes: DMC \cite{dmc}, MXT \cite{mxt}, TMCC \cite{tmcc}, and DyLeCT \cite{dylect}. For DMC, promotion to line-level compression occurs for each touch on data compressed in block-level compression, whereas demotion occurs on unaccessed data every 50 million cycles in the background (as in \cite{dmc}).
Regarding MXT, it mitigates the overhead of block-level compression by defining a \textit{caching region} (i.e., the promoted region in IBEX) indexed by an on-chip SRAM tag array, where its latency values are obtained with CACTI 7.0 \cite{cacti}.
TMCC employs zsmalloc-like allocator to support variable-size chunks. We evaluate its base system without the page table modification so the design remains deployable within CXL memory.
DyLeCT builds on the TMCC base system but additionally maintains two versions of metadata tables: a pre-gathered table, which stores short translation entries for the most frequently accessed pages, and a unified table, which holds normal entries for the rest.
We also compare against Compresso \cite{compresso}, a representative line-level compression that offers high performance compared to block-level compression despite a lower compression ratio.

\begin{table}[h]
\centering
\caption{Workloads for evaluation} \label{tab:workload}
\begin{tabular}{|p{0.25\linewidth}|p{0.21\linewidth}|p{0.18\linewidth}|p{0.18\linewidth}|}
\hline
\textbf{Benchmark} & \textbf{Workloads} & \textbf{RPKI} & \textbf{WPKI} \\ 
\hline
\multirow{5}{*}{CPU2017 \cite{cpu2017}} & bwaves  & 13.4 & 2.1 \\ 
\cline{2-4}
                   & mcf      & 55.0 & 9.6 \\ 
\cline{2-4}
                   & parest   & 14.5 & 0.2 \\ 
\cline{2-4}
                   & lbm      & 23.9 & 17.8 \\ 
\cline{2-4}
                   & omnetpp  & 8.8  & 4.1 \\ 
\hline
\multirow{4}{*}{GAPBS \cite{gapbs}} & bfs  & 41.9  & 2.7 \\ 
\cline{2-4}
                   & pr       & 126.8 & 2.3 \\ 
\cline{2-4}
                   & cc       & 33.3  & 3.8 \\ 
\cline{2-4}
                   & tc       & 16.7  & 11.6 \\ 
\hline
XSBench \cite{xsbench} & XSBench & 37.7 & 0.0 \\ 
\hline
\end{tabular}
\end{table}

\noindent\textbf{Workloads:} 
Table~\ref{tab:workload} shows our evaluated workloads, along with their memory reads and writes per kilo-instruction (RPKI \& WPKI): (1) various memory-intensity condition is tested with five workloads of CPU2017 \cite{cpu2017}; (2) pointer-chasing access behaviors are tested with five graph kernels from GAPBS \cite{gapbs} using Twitter \cite{graph-twitter} as the input graph; (3) highly irregular access behavior is also tested with XSBench that is used for Monte Carlo-based scientific simulation.
Workloads are run in a multi-programmed fashion, where process identifiers are assigned to prevent data sharing between the same virtual address but different processes.

\noindent\textbf{Fast-forwarding:} 
Simulating memory-intensive workloads on SST takes a long simulation time. For instance, simulating a memory-intensive \textit{pr} requires more than 13 hours to simulate a single-core system running 5 billion instructions on a server. Hence, we first fast-forward all workloads to their representative execution intervals and then simulate for 1 billion instructions. 
Fast-forwarding is achieved in two ways to evaluate different types of workloads.
For SPEC CPU2017 workloads, we use SimPoint \cite{simpoint} to identify target intervals that exclude the initialization phase.
For GAPBS and XSBench, we skip the input-loading phase to exclude one-time overheads. After the input-loading phase, we also bypass an additional 1 billion instructions to \textit{warm up} the stack segment, because the zero-initialized variables could overstate compression ratios. 
For all workloads, simulation starts in the cold state (i.e., an empty promoted region and caches in the memory hierarchy) after fast-forwarding, with 13 million instructions sufficient to fill the LLC.

\section{Evaluation Results} \label{sec:eval-results}
\subsection{Comparison of Different Schemes} \label{sec:eval-results:comparison}

For the comparison of different schemes, we evaluate the efficiency using two critical metrics of memory compression schemes: normalized performance (defined here as \textit{the inverse of execution time}) and compression ratio. In this subsection, IBEX incorporates all optimization schemes, including shadowed promotion, block co-location, and metadata compaction. The metadata cache size is kept constant across all schemes to ensure a fair comparison.

\begin{figure}[h]
\centering
\includegraphics[scale=0.75]{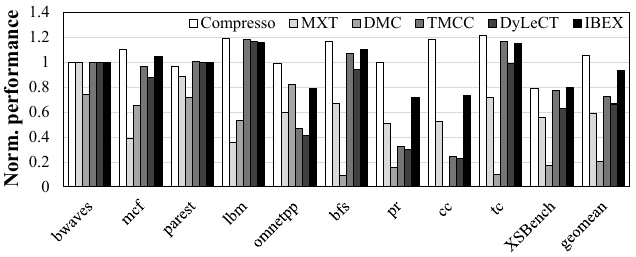}
\caption{Normalized performance of different schemes (512MB caching region for MXT, TMCC, DyLeCT and IBEX).} \label{fig:performance_comparison}
\end{figure}

\noindent\textbf{Performance:}
Figure~\ref{fig:performance_comparison} shows the performance of each scheme relative to uncompressed CXL memory.
Compresso, a line-level compression scheme, achieves the highest performance among the evaluated schemes due to its lightweight management overhead. However, this performance advantage comes at the expense of a lower compression ratio, as discussed further in the subsequent compression-ratio analysis.
Among the block-level schemes, IBEX delivers the highest performance, achieving average speedups of 1.28$\times$ over TMCC and 1.40$\times$ over DyLeCT. TMCC’s variable‑size chunks increase translation coverage but lose much of their benefit when the spatial locality of OS-level pages is weak across page boundaries, while DyLeCT further reduces TMCC’s performance by requiring two metadata‑table probes on a miss.
MXT and DMC show significantly lower performance, with IBEX outperforming them by 1.58$\times$ and 4.64$\times$, respectively. 
%
%
DMC exhibits the lowest performance because the limited internal bandwidth of CXL memory is insufficient to handle the large 32KB transfers.
Notably, IBEX shows performance improvements for workloads that frequently access zero pages, including \textit{lbm}, \textit{bfs}, and \textit{tc}, because the use of metadata \texttt{type} bits eliminates any memory accesses for zero pages on a metadata cache hit.

In general, block-level schemes exhibit notable performance degradation for some workloads, such as \textit{omnetpp}, \textit{pr}, and \textit{cc}. The degradation stems from an insufficient promoted region size, resulting in repeated page promotions and demotions. IBEX mitigates this penalty more effectively than its counterparts by reducing unnecessary recompression through shadowed promotion and enabling fine-grained promotion via block co-location. Furthermore, this performance loss can be alleviated by configuring a larger promoted region at boot time; we observe that allocating a 1GB promoted region reduces the degradation to 3\% for these workloads.
%

\begin{figure}[h]
\centering
\includegraphics[scale=0.75]{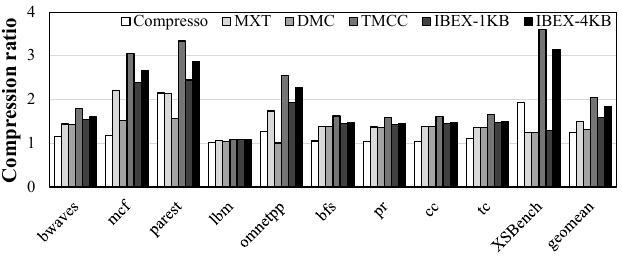}
\caption{Compression ratios regarding different schemes.} \label{fig:compression_ratio}
\end{figure}

\noindent\textbf{Compression ratio:} 
We evaluate compression ratios over the \textit{entire} execution. Figure~\ref{fig:compression_ratio} presents the compression ratios of various schemes, computed as the geometric mean of ratios sampled every 100 billion instructions throughout execution.
To conservatively assess the capacity benefits, unaccessed memory regions (i.e., filled with zero) are excluded when calculating the compression ratio. Furthermore, we evaluate IBEX with two block sizes to analyze the impact of block size on compression ratio, namely IBEX-4KB and IBEX-1KB. DyLeCT, which shares the same base system as TMCC, is excluded from this comparison.
With a block size of 1KB, IBEX achieves an average compression ratio of 1.59, outperforming MXT (1.49), thanks to its finer-grained chunk allocation. Compresso, operating at the line-level, yields the lowest compression ratio of 1.24, which is \textbf{22\%} lower than IBEX-1KB. Its capacity benefits are limited to a narrow subset of workloads, providing less than a 5\% capacity improvement for \textit{lbm} and graph workloads. Since securing sufficient memory capacity is critical to prevent out-of-memory failures and thereby high system availability, the marginal capacity gains of line-level compression limit its practicality in such hyperscale environments.
DMC, which combines line- and block-level compression, achieves a moderate compression ratio of 1.31.
TMCC’s capacity gains come at the cost of managing 64 possible variable-size chunk configurations, necessitating sophisticated management.

\subsection{Breakdown Analysis of IBEX} \label{sec:eval-results:breakdown}

\begin{figure}[h]
\centering
\includegraphics[scale=0.68]{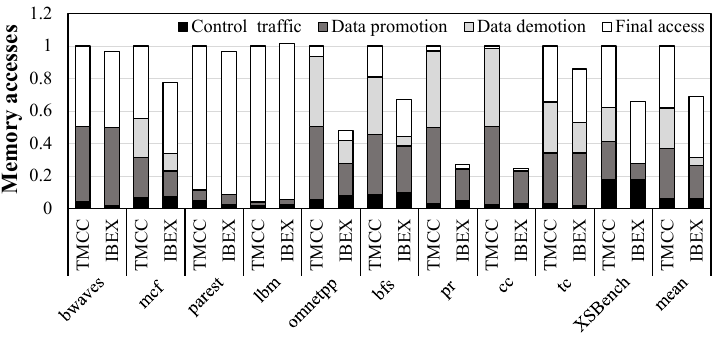}
\caption{Memory access breakdown in TMCC and IBEX, where the access of each workload is normalized to TMCC.} \label{fig:tmcc_vs_ibex}
\end{figure}

\noindent\textbf{Reduction in demotion overhead:}
IBEX is a complete hardware compression mechanism tailored for CXL memory, wherein internal bandwidth is limited due to form-factor constraints. 
Figure~\ref{fig:tmcc_vs_ibex} breaks down the memory accesses to comprehensively understand the performance benefits of IBEX over the state-of-the-art scheme, namely TMCC. Please note that \textit{control traffic} refers to memory accesses incurred by metadata management and recency tracking, while \textit{final access} denotes the number of memory accesses required to retrieve the data in the promoted region. 
It is noteworthy that \textit{bwaves}, \textit{parest}, \textit{lbm}, and \textit{XSBench} do not exhibit demotion traffic. In particular, the promoted regions of \textit{bwaves}, \textit{parest}, and \textit{lbm} are large enough to retain promoted data without eviction. On the other hand, the read-only attribute of \textit{XSBench} results in zero-demotion traffic in IBEX, owing to the shadowed promotion scheme. 

As shown in Figure~\ref{fig:tmcc_vs_ibex}, IBEX incurs 30\% less total memory traffic on average than TMCC. This difference is particularly evident in workloads with frequent page migrations. For example, in \textit{pr} and \textit{cc}, promotion and demotion together account for 93\% and 95\% of TMCC’s total memory accesses, respectively. Since IBEX avoids a substantial portion of this internal data movement, its total memory accesses are nearly three-quarters lower than TMCC’s in both workloads (72\% lower in \textit{pr} and 75\% lower in \textit{cc}).
We observe that these savings arise from two mechanisms. First, IBEX leverages block co-location to enable fine-grained promotion at the 1KB block granularity rather than the page granularity. As a result, IBEX achieves 34\% less promotion traffic than TMCC. Second, shadowed promotion allows skipping the recompression process of unmodified promoted data. The benefit of skipping recompression becomes pronounced for read-intensive workloads such as \textit{pr}, \textit{cc}, and \textit{XSBench}, where IBEX reduces demotion traffic by more than 99\% relative to TMCC.

\begin{figure}[h]
\centering
\includegraphics[scale=0.75]{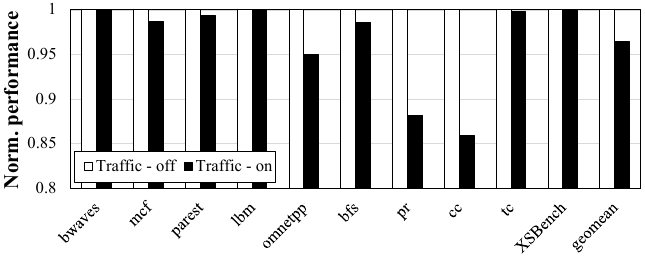}
\caption{Normalized performance of IBEX with (practical) and without (miracle) background traffic modeling.} \label{fig:breakdown_demotion}
\end{figure}

\noindent\textbf{Impact of demotion candidate scanning:}
IBEX employs a demotion engine running in the background to identify cold pages. This process generates two types of background
traffic: (1) requests to update the reference bits of recently accessed pages, and (2) requests to scan for the demotion candidate when the promoted region runs out of space.
Figure~\ref{fig:breakdown_demotion} quantifies the performance impact of such traffic by comparing the performance of IBEX to an idealized version in which the traffic overhead is excluded. 
The figure demonstrates that the performance degradation is minimal in most cases; that is, 1\% slowdown is observed for \textit{mcf} and \textit{parest}. 
However, the frequency of page demotions increases significantly when the promoted region is insufficient for some workloads. This increases background traffic as the demotion engine actively scans the activity region, resulting in non-negligible performance degradation. In such cases, we observe that background traffic causes 5\% slowdown for \textit{omnetpp} and 13\% slowdown for \textit{pr} and \textit{cc}.

\begin{figure}[h]
\centering
\includegraphics[scale=0.75]{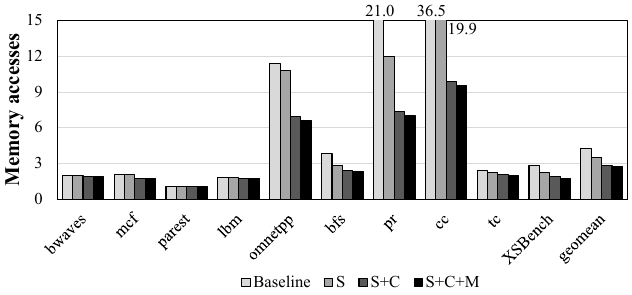}
\caption{Traffic breakdown of IBEX.} \label{fig:breakdown_accesses}
\end{figure}

\noindent\textbf{Traffic reduction by each optimization:}
We incrementally apply shadowed promotion (denoted as `S'), co-location scheme (denoted as `C'), and metadata compaction (denoted as `M') to the baseline IBEX. For the baseline and the `S'-only, the latency of both compression and decompression is configured 4$\times$ larger than the values in Table~\ref{tab:sys-config}, as the block size is 4KB for these two scenarios.
Figure~\ref{fig:breakdown_accesses} presents the number of memory accesses, normalized to the memory access count of an uncompressed system.
On average, shadowed promotion, co-location of blocks, and metadata compaction reduce the memory accesses by 16\%, 20\%, and 3.3\%, respectively.
The impact of optimizations becomes more significant when frequent page promotions and demotions occur. For \textit{omnetpp}, \textit{pr}, and \textit{cc}, the undersized promoted region causes baseline IBEX to incur 20.6$\times$ more memory accesses on average compared to uncompressed memory. In this case, shadowed promotion reduces traffic by 34\%, and subsequent block co-location further cuts remaining memory accesses by 42\%.

\subsection{Sensitivity Analysis} \label{sec:eval-results:sensitivity}

\begin{figure}[h]
\centering
\includegraphics[scale=0.75]{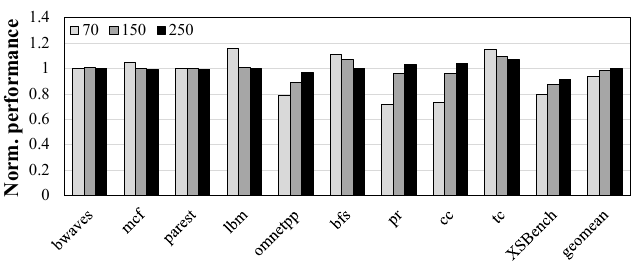}
\caption{Normalized performance relative to uncompressed memory system regarding different CXL latency values.} \label{fig:sensitivity_cxl}
\end{figure}

\noindent\textbf{Sensitivity to CXL latency:}
Figure~\ref{fig:sensitivity_cxl} shows that IBEX's performance converges to that of the uncompressed system as CXL latency increases. 
For \textit{lbm}, \textit{bfs}, and \textit{tc}, performance improvements are observed under the 70ns configuration due to frequent zero-page accesses. As CXL latency increases, the benefit of skipping zero-page accesses diminishes. In contrast, \textit{omnetpp}, \textit{pr}, \textit{cc}, and \textit{XSBench} experience performance degradation compared to uncompressed memory under the 70ns configuration. At higher CXL latencies, the impact of compression overhead is less pronounced, resulting in a smaller degree of performance degradation.
Notably, \textit{pr} and \textit{cc} (i.e., high MPKI) exhibit the most significant variations in relative performance as CXL latency increases. \textit{Interestingly}, higher CXL latency causes outstanding requests to occupy hardware longer (e.g., host’s MSHRs); hence, it decreases the rate at which new memory requests can be issued. Consequently, the system shifts toward latency-bounded, lowering the internal bandwidth load and alleviating performance degradation associated with congestion.
%

\begin{figure}[h]
\centering
\includegraphics[scale=0.75]{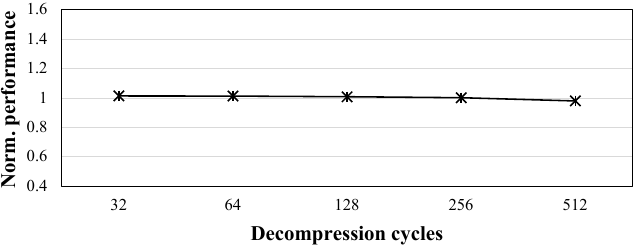}
\caption{Performance sensitivity to decompression cycles.} \label{fig:sensitivity_decompression}
\end{figure}

\noindent\textbf{Decompression latency:}
Figure~\ref{fig:sensitivity_decompression} illustrates IBEX's average performance across different decompression cycles relative to the uncompressed system. The size of the promoted region is set to 1024MB to eliminate the effects of allocating an undersized promoted region.
As shown in the figure, IBEX maintains performance comparable to the uncompressed system. Although increasing decompression latency incurs a slight performance degradation, the overall impact is minimal---a mere 2\% drop relative to the uncompressed system. These observations demonstrate IBEX’s \textit{robustness} against larger decompression cycles, enabling the integration of more sophisticated block-level compression algorithms to achieve higher compression ratios with minimal performance trade-offs.

\begin{figure}[h]
\centering
\includegraphics[scale=0.75]{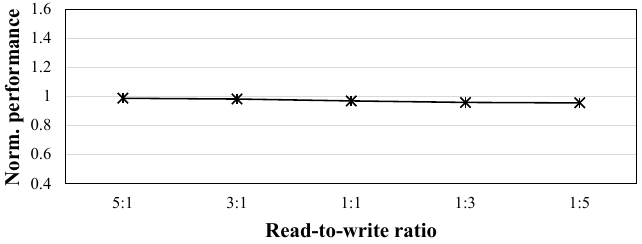}
\caption{Performance sensitivity to write intensity.} \label{fig:sensitivity_readtowrite}
\end{figure}

\noindent\textbf{Write intensity:}
We evaluate IBEX’s performance under varying levels of write intensity by instrumenting the memory requests of \textit{XSBench}---a benchmark that exhibits 100\% read requests---into write requests using binomial probabilities to generate five read-to-write ratios between 5:1 and 1:5. 
Figure~\ref{fig:sensitivity_readtowrite} illustrates IBEX’s performance across these ratios, normalized to the read-only baseline. As write intensity increases, IBEX experiences a minor slowdown, peaking at around 4\%, because write intensity may reduce the benefits of shadow promotion. Still, IBEX also exhibits solid robustness even under a high write-intensity scenario.

\section{Discussion}
\label{sec:discussion}
\noindent\textbf{Scalability to larger systems:}
While our evaluation results are limited to a 4-core configuration due to the high cost of cycle-accurate simulation for memory-intensive workloads, this setup does not fully represent the larger-scale environments in which CXL memory is likely being deployed. Hyperscale systems typically employ substantially higher core counts, often in multi-socket configurations with tens to hundreds of cores. As systems scale, increased concurrency is likely to amplify both data traffic and compression-related traffic, placing additional pressure on the internal bandwidth of CXL memory expanders with limited channel counts. Accordingly, the advantages of IBEX are likely to be further pronounced in such environments, as its design is architecturally more resilient to such bandwidth limitations by explicitly minimizing internal bandwidth overhead.
\noindent\textbf{Implications in page fault rates:}
Due to the capability limits of the SST simulator, the evaluation methodology assumes that memory compression can perfectly prevent page swaps (i.e., major page faults) between system memory and storage and lacks system-level performance quantification. Nonetheless, we have tried to evaluate the reduction in page fault rates by modeling a simple LRU-based list that records pages in use in our simulator. We count the number of replacements and compare two scenarios: (1) an uncompressed baseline system where physical memory capacity is restricted to 50\% of the workload's working set, and (2) an IBEX-enabled system with the same physical memory, which achieves a larger effective capacity through memory compression.
%

\begin{figure}[h]
\centering
\includegraphics[scale=0.75]{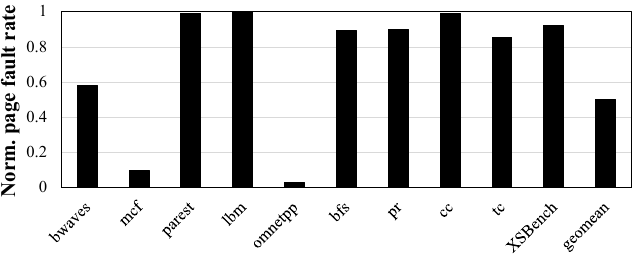}
\caption{Normalized page fault rates relative to the uncompressed memory under memory-constrained conditions.} \label{fig:pagefault_rate}
\end{figure}

Figure~\ref{fig:pagefault_rate} presents the normalized page fault rates of IBEX relative to the uncompressed baseline. The results demonstrate that the effective capacity expansion provided by IBEX allows a larger portion of the working set to reside in main memory, yielding an average 49\% reduction in page faults. Notably, this reduction is highly workload-dependent and correlates closely with data compressibility. For instance, highly compressible workloads such as \textit{omnetpp} and \textit{mcf}, exhibit substantial page fault reductions of 90\% and 97\%, respectively. Conversely, for workloads with inherently low compressibility like \textit{lbm}, the system naturally exhibits fault rates nearly identical to the uncompressed baseline.
Notably, a high compression ratio does not universally guarantee a significant reduction in page faults, as observed in \textit{parest}. Despite exhibiting a high compression ratio, its fault reduction remains marginal (0.8\%). This is because 99\% of its baseline page faults are cold faults incurred on first data access, rather than faults caused by memory pressure. 
Overall, these preliminary observations indicate that the capacity benefits from IBEX effectively translate into system-level performance benefits by alleviating storage I/O bottlenecks in capacity-constrained environments.

\section{Related Work}
\label{sec:relatedWork}

\noindent\textbf{Hardware-based memory compression:} 
Previously, several \textit{line-level} compression schemes have been proposed for practical systems \cite{rmc,lcp,buddy,compresso,buri}. Despite several optimizations on metadata management, these line-level approaches offer limited improvements in memory capacity. On the other hand, \textit{block-level} compression significantly enhances effective memory capacity but demands careful management of associated performance overhead. Therefore, prior works selectively apply block-level compression to main memory, based on the recency of each block \cite{mxt,dmc,tmcc,dylect}.

For example, MXT \cite{mxt} mitigates performance overhead using a 32MB caching region indexed by an on-chip tag array, which is insufficient for today's memory-intensive workloads. Adapting MXT to modern systems would require substantial on-chip resources to address the larger cache, and hence, scalability is limited.
DMC \cite{dmc} combines line- and block-level compression to balance capacity and performance, simplifying metadata management by migrating data in coarse 32KB chunks. Such large transfers are practical only with high-bandwidth memory (e.g., HMC) that can hide the migration overhead.
TMCC \cite{tmcc} aims to reduce translation latency: it compresses each page table block and reuses the freed space to store translation metadata, so that a page table walk implicitly prefetches the required mappings. However, realizing TMCC requires modifications to the CPU-side memory controller, caches, and MMU, whereas IBEX can be realized within the CXL device.
DyLeCT \cite{dylect} employs short metadata entries for frequently accessed pages using set-associative placement, achieving huge-page-like translation reach by aggregating short entries. However, it requires probing both short and normal metadata on cache misses, incurring additional traffic in internal bandwidth-constrained systems.

Since hardware-only memory compression decouples OS-level physical memory from actual memory size, imprecise memory allocation can cause performance interference, mainly due to potential data spilling to storage. Therefore, DMU \cite{dmu} allows the OS to specify a per-job objective for actual DRAM usage. However, this mechanism depends on explicit coordination with the OS and cannot be implemented solely at the device level.

\noindent\textbf{Software-defined far memory:} 
Several works \cite{sdfm-in-wc, TMO, XFM} have explored software-controlled paging mechanisms and compression to optimize memory usage. These techniques are built on top of Linux’s \textit{zswap}, creating a software-defined far memory (SDFM) where cold pages are compressed and stored to increase effective capacity. Some works \cite{sdfm-in-wc, TMO} incorporate a software-level profiler to manage compression while accounting for its impact on performance. However, these software-driven approaches rely on the CPU for compression and decompression, inevitably consuming CPU cycles and interface bandwidth.
XFM \cite{XFM} addresses this limitation by integrating a near-memory accelerator that offloads compression. Nevertheless, these approaches require kernel-level modifications beyond the memory module itself, posing challenges for memory vendors who seek self-contained solutions.

\section{Conclusion}
\label{sec:conclusion}
The physical constraints of CXL memory fundamentally impede scalable memory expansion in hyperscale systems. To address this problem, this paper presents a scalable compression architecture for CXL memory, IBEX. Our solution minimizes the management overhead of compressed blocks, enhancing effective memory capacity without incurring notable performance degradation. IBEX also encompasses shadowed promotion that significantly mitigates demotion penalties. Also, IBEX includes several metadata downsizing techniques to further reduce metadata access overhead. 
Consequently, IBEX achieves 1.28$\times$--1.40$\times$ speedups over previous promotion-based block-level schemes. 
%


\begin{acks}
We appreciate all the valuable comments from the anonymous reviewers of ICS 2026, HPCA 2026, MICRO 2025, and ISCA 2025. 
This work was supported 
in part by K-CHIPS (Korea Collaborative \& High-tech Initiative for Prospective Semiconductor Research) (2410012307, RS-2025-02305531, 25063-15TC) funded by the Ministry of Trade, Industry \& Energy (MOTIE, Korea), 
in part by Institute for Information \& communications Technology Planning \& Evaluation (IITP) grant funded by the Korea government (MSIT) (RS-2025-02314443). 
Hyokeun Lee is the corresponding author. 

\end{acks}

\bibliographystyle{ACM-Reference-Format}
\bibliography{refs}

\end{document}